\documentclass[aps,prd,preprint,superscriptaddress,nofootinbib]{revtex4}

\usepackage{epsfig,verbatim}
\usepackage{subfigure}
\usepackage{amsmath, amssymb, graphics}
\usepackage{color}
\usepackage{slashed}            
\usepackage{bbm}                
\usepackage{units}              
\usepackage{xspace}             
\usepackage{enumerate}          
\usepackage{soul}
\usepackage{epstopdf}
\usepackage{tensor}
\usepackage{comment}
\usepackage[utf8]{inputenc}

\newcommand{\mathsym}[1]{{}}

\renewcommand\({\left(}
\renewcommand\){\right)}
\renewcommand\[{\left[}
\renewcommand\]{\right]}

\newcommand{\ba}{\begin{eqnarray}}
\newcommand{\ea}{\end{eqnarray}}
\newcommand{\be}{\begin{equation}}
\newcommand{\ee}{\end{equation}}
\newcommand{\LL}{\mathcal{L}}
\newcommand{\OO}{\mathcal{O}}

\newcommand{\RR}{\mathcal{R}}

\begin{document}




\title{The two-field regime of natural inflation}

\author{Ana Ach\'ucarro}
\affiliation{Instituut-Lorentz for Theoretical Physics, Universiteit Leiden, 2333 CA Leiden, The Netherlands}
\affiliation{Department of Theoretical Physics, University of the Basque Country, 48080 Bilbao, Spain}
\author{Vicente Atal}
\email[]{atal@lorentz.leidenuniv.nl}
\affiliation{Instituut-Lorentz for Theoretical Physics, Universiteit Leiden, 2333 CA Leiden, The Netherlands}
\author{Masahiro Kawasaki}
\affiliation{Institute for Cosmic Ray Research, The University of Tokyo, 5-1-5 Kashiwanoha,Kashiwa, Chiba 277-8582, Japan}
\affiliation{Kavli IPMU (WPI), UTIAS, The University of Tokyo, 5-1-5 Kashiwanoha, Kashiwa, 277-8583, Japan}
\author{Fuminobu Takahashi}
\affiliation{Kavli IPMU (WPI), UTIAS, The University of Tokyo, 5-1-5 Kashiwanoha, Kashiwa, 277-8583, Japan}
\affiliation{Department of Physics, Tohoku University, Sendai 980-8578, Japan}





\date{\today}
  


\begin{abstract}
The simplest two-field completion of natural inflation has a regime in which both fields are active and in which its predictions are within the Planck 1-$\sigma$ confidence contour. We show this for the original model of natural inflation, in which inflation is achieved through the explicit breaking of a U(1) symmetry. We consider the case in which the mass coming from explicit breaking of this symmetry is comparable to that from spontaneous breaking, which we show is consistent with a hierarchy between the corresponding energy scales. While both masses are comparable when the observable modes left the horizon, the mass hierarchy is restored in the last e-foldings of inflation, rendering the predictions consistent with the isocurvature bounds. For completeness, we also study the predictions for the case in which there is a large hierarchy of masses and an initial period of inflation driven by the (heavy) radial field.
\end{abstract}

\maketitle





\section{Introduction}
One of the most prominent models of inflation is natural inflation~\cite{Freese:1990rb}. In its original version, the inflaton is the Nambu-Goldstone boson of a spontaneously broken U(1) symmetry.   In particular, the inflaton is the phase of a complex field whose modulus is strongly stabilized. A mass for the inflaton, stable under radiative corrections, can be generated via explicit symmetry breaking terms. For example, instanton effects can create such a potential, as in the case of the Peccei-Quinn mechanism for strong CP conservation~\cite{Peccei:1977hh,Peccei:1977ur,Weinberg:1977ma,Wilczek:1977pj}. The U(1) symmetry is broken to a discrete subgroup $\phi \rightarrow \phi+2\pi f$, and the potential for the inflaton (in this case an axion) is given by:
\be\label{eq:potvai}
V=\Lambda^4\(1+\cos(\phi/f)\) \ ,
\ee
where $f$ is the so-called axion decay constant, $\Lambda$ is a dynamically generated scale that sets the overall magnitude of the potential and a cosmological constant has been tuned to make the potential vanish at the minimum of the potential. While this model is under good theoretical control for subplanckian values of $f$ (where the Planck mass $M_{\text{pl}}=1.22\times10^{19}$GeV), recent data by the Planck satellite~\cite{Ade:2015lrj} seems not to be consistent with its predictions for the tilt of the power spectrum $n_s$, and tensor to scalar ratio $r$. Furthermore, the region in which the tension between the prediction and the observations is less severe is for super planckian values of $f$, in which one may expect the low energy effective theory to break down~\cite{Banks:2003sx}\footnote{However, several mechanisms to achieve the potential (\ref{eq:potvai}) with $f>M_{\text{pl}}$ consistent with a low energy description have been put forward in the literature~\cite{Kim:2004rp,Dine:2014hwa,Yonekura:2014oja, Harigaya:2014eta}. It has also been pointed out that non-perturbative dynamics in the single field potential can make the predictions for subplanckian values of $f$ consistent with CMB data~\cite{Albrecht:2014sea}.}. This has motivated the study of modifications of the single field potential (\ref{eq:potvai}), in order to test if the tension persists with theoretically and/or phenomenologically well motivated extensions. For example, the predictions in the $(n_s,r)$ plane might be substantially affected when considering  many axions~\cite{Dimopoulos:2005ac,Czerny:2014wza,Peloso:2015dsa}, multiple sinusoidal functions~\cite{Czerny:2014wza,Czerny:2014xja}, extra non-renormalizable operators~\cite{Croon:2014dma,McDonald:2014nqa}, different periodic functions~\cite{Higaki:2015kta}, and/or higher dimensional theories \cite{Neupane:2014vwa}. Some  completions are based on the fact that the constants that determine the low energy potentials are, in general, vacuum expectation values (v.e.v) of additional 
fields, and that their dynamics can be non-trivial. For example, even if these additional fields are very heavy with respect to the energy scale of inflation, they can induce changes in the effective potential \cite{Kappl:2015pxa,Dudas:2015lga,Li:2015taa} and/or in the speed of sound of the inflaton fluctuations \cite{Achucarro:2015rfa}. The Nambu-Goldstone boson may also take different functional forms, depending on the symmetry that is being broken~\cite{Cohn:2000hc,Ross:2009hg,Burgess:2014tja,Burgess:2014oma,Croon:2015fza}, or, e.g. the field content in higher dimensional theories~\cite{Kaloper:2008fb,McAllister:2008hb}. 

We aim to explore the multifield regime of natural inflation in a very simple completion in which, in addition to the angular field $\theta= {\rm arg}[\Phi]$ (which has a sinusoidal potential), there is a radial field $r=|\Phi|$ which is not strongly stabilized. We will work with \emph{same} original model as proposed in \cite{Freese:1990rb}, in which there is a quartic spontaneous symmetry breaking potential together with a term $V \propto \Phi+\bar{\Phi}$ that explicitly breaks the U(1) symmetry.  Irrespective of the hierarchy between the explicit and spontaneous symmetry breaking scales, the masses of both radial and angular field may or may not be comparable. On the one hand, if the radial field is very heavy with respect to the scale of inflation, its v.e.v will be strongly stabilized and determined only by the spontaneous symmetry breaking scale. The effective potential will then reduce to (\ref{eq:potvai}). This case is well understood, and its predictions were first computed in~\cite{Freese:1990rb}. On the other hand, if both masses are similar none of the fields will be strongly stabilized, and the dynamics of both of them will become important. The objective of this paper is to study 
this latter case. 

\section{Natural models}

We will study the following Lagrangian:
\be\label{eq:pot}
\LL=\frac{1}{2}\partial_{\mu}\Phi \partial ^{\mu}\bar{\Phi}-\lambda\(r_0^2-|\Phi|^2\)^2-\Lambda^{3}\(\Phi+\bar{\Phi}\)\ ,			
\ee
\noindent which is a very simple completion of (\ref{eq:potvai}). Here $r_0$ is the scale of the spontaneous $U(1)$ breaking and $\Lambda$
represents the scale of the explicit breaking. As we will see later, generating the right amplitude for the two-point function of the curvature pertubations fixes $r_0 \gg \Lambda$. Writing $\Phi=r\ e^{i\theta}$, the Lagrangian can be written as
\be\label{eq:lag2}
\LL=\frac{1}{2}\partial_{\mu}r\partial^{\mu}r+\frac{1}{2}r^2\partial_{\mu}{\theta}\partial^{\mu}{\theta}-\lambda\(r_0^2-r^2\)^2-2\Lambda^{3}r \cos \theta \ . 
\ee
It is useful to define the following dimensionless quantity
\be
\beta=\frac{2\Lambda^{3}}{r_0^{3}\lambda} \ ,
\ee
so the potential can be written as
\be
V=\mu^4\[\(1-\(\frac{r}{r_0}\)^2\)^2+\beta\(\frac{r}{r_0}\)\cos \theta\]
\ee
\noindent where $\mu^4=\lambda r_0^4$. Written in this way, it is clear that the only parameters that determine the dynamics of the theory are $r_0$ and $\beta$, since $\mu$ is an overall factor which is fixed by the amplitude of the two-point function\footnote{Furthermore, $r_0$ can not be absorbed into the definition of $r$ since the kinetic term only depends on $r$ and not on the combination $r/r_0$.}. While both $r_0$ and $\beta$ will determine the axion decay constant at low energies (equal to $r_0$ when $\beta\ll 1$), only $\beta$ controls the hierarchy between the masses of the radial and angular field. Importantly,  a value $\beta\sim\OO(1)$ will still imply a hierarchy between $\Lambda$ and $r_0$, of the order $\Lambda/r_0\sim 10^{-4}$, as a rather small value for $\mu$ (which implies a small value for $\lambda$ when $r_0\sim M_{\text{pl}}$) is needed in order to fix the amplitude for the two-point function. In other words, multifield dynamics ($\beta\sim\OO(1)$) will still imply a hierarchy 
between both energy scales. As corrections to the potential are expected to go as $(\Lambda/r_0)^n$  -where $n$ is some positive power- higher order contribautions will still be under control.\footnote{Take for example the case in which $\Lambda$ is generated by non-perturbative gravitational effects~\cite{Kallosh:1995hi}. In this case $\Lambda^3=e^{-S}M_{\text{pl}}^3$, where $S$ is the action a wormhole configuration. Contributions of the order $e^{-nS}\cos(n\theta)$ are going to be $(\Lambda^3/M_{\text{pl}}^3)^n\cos(n\theta)$ which are small considering that $\Lambda/r_0\ll 1$ and $r_0\sim M_{\text{pl}}$.} 

Defining $x=r/r_0$ and adding a cosmological constant $V_0$, we write the potential as
\be
\frac{V}{\mu^4}=\(1-x^2\)^2+\beta x\cos\theta+\frac{V_0}{\mu^4}\ .
\ee
Minimizing in the radial direction at $\theta=\pi$, we get
\be
\frac{dV}{dr}\Big|_{\theta=\pi}=0 \quad\quad \rightarrow\quad\quad 4\(1-x^2\)x+\beta=0 \ . \label{eq:min}
\ee
Evaluating the potential at $x$ satisfying (\ref{eq:min}) gives the contribution of the cosmological constant. As the second term in (\ref{eq:min}) is non-zero, the absolute minimum is reached for values of $x\neq$ 1 (or equivalently $r\neq r_0$). The deviation from $r=r_0$ is going to be negligible when $\beta\ll1$ but becomes important for values of $\beta\sim1$. Since in the minimum of the potential the field is stabilized at a radius $r\neq r_0$, the low energy effective theory will have a decay constant that is in general different from $r_0$.\\

\begin{figure}[ht!]
  \includegraphics[width=300pt]{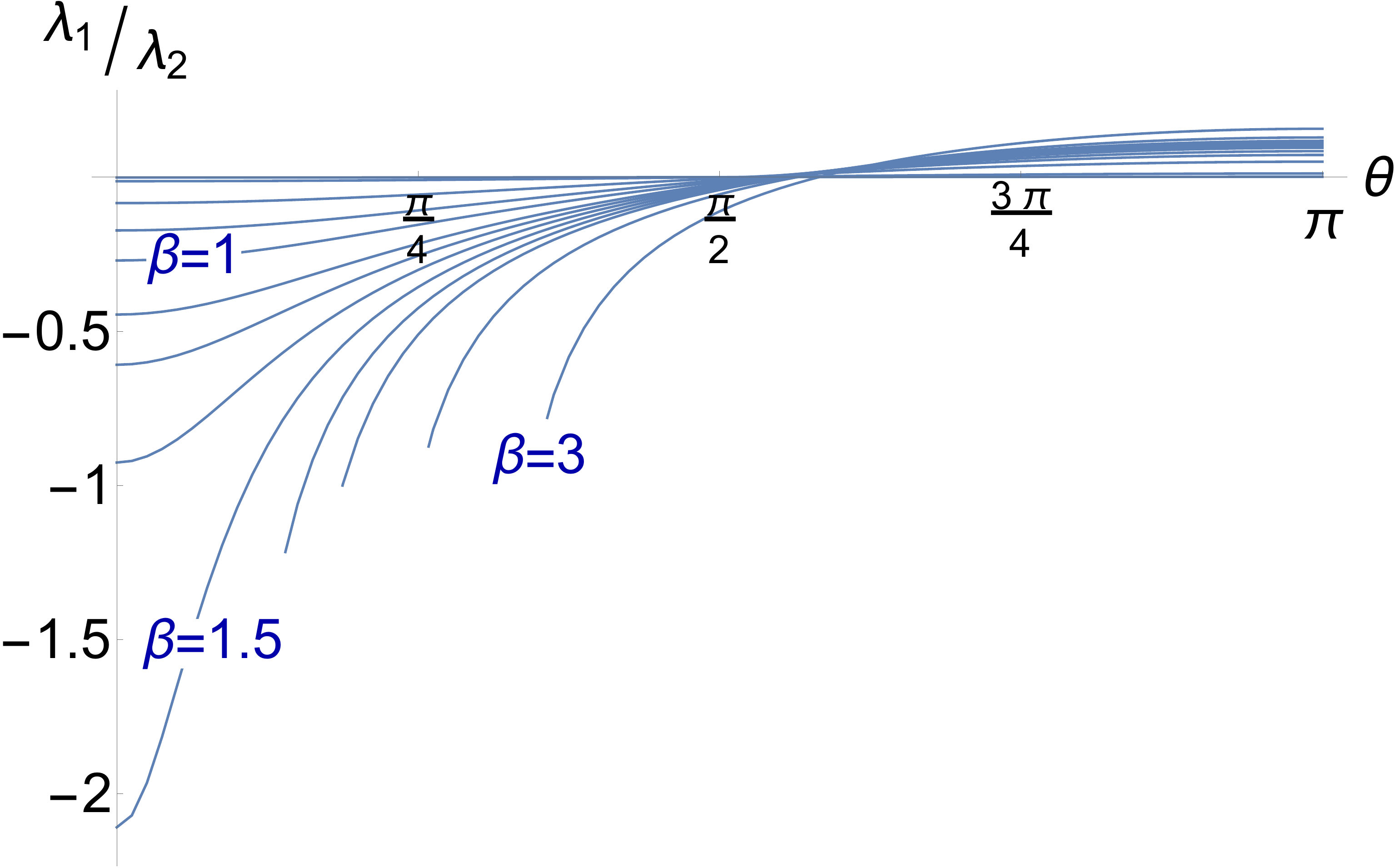}
\caption{Ratio of the two eigenvalues of the mass matrix evaluated at the radial minimum, for different values of $\beta$.}
\label{fig:eig}
\end{figure}

One necessary condition for having a single field description is that at the instantaneous minimum the mass of the orthogonal field is much bigger than the Hubble scale $H$.\footnote{A single field description does not necessarily mean that the heavy field can be \emph{truncated}. Even if there is no particle production, a heavy field may a have a time-dependent v.e.v and/or be displaced from the minimum, modifying the low energy effective potential~\cite{Rubin:2001in,Dong:2010in,Buchmuller:2015oma} and/or speed of sound of the fluctuations \cite{Achucarro:2010da}.} We can thus compute the mass eigenvalues at the minimum, and see how they change as we vary $\beta$. With canonical kinetic terms (as given in eq. (\ref{eq:lag2})), the mass matrix is given by
\be
V^I_{\phantom{I}J} =\frac{\mu^4}{r_0^2}
 \begin{pmatrix}
  12x^2-4 & -\frac{\beta\sin \theta}{x} \\
  -\frac{\beta  \sin \theta}{x} & -\frac{\beta\cos \theta}{x}  \\
  \end{pmatrix}  \ ,
\ee
\noindent and the eigenvalues are then:
\be
\lambda_{1,2}=\frac{\mu^4}{r_0^2}\[6x^2-2-\frac{\beta\cos \theta}{2x}\pm \[\(6x^2-2-\frac{\beta\cos \theta}{2x}\)^2-\Delta\]^{1/2}\]
\ee
with $\Delta$ the determinant of the mass matrix (ignoring the prefactor $\mu^4/r_0^2$). In the limit $\beta\ll 1 $, $\lambda_1$ and $\lambda_2$ correspond to the mass (squared) eigenvalues of $\theta$ and $r$ respectively. When $\beta \sim \OO(1)$, the mass eigenstates are a linear combination of the radial and angular field. In figure \ref{fig:eig} we evaluate the ratio $\lambda_1/\lambda_2$  at the radial instantaneous minimum of the potential, as a function of $\beta$. 

We can see that for values of $\beta <1$ there is a sufficient hierarchy in the mass eigenvalues at the minimum of the potential. Provided that there are no strong turns in the trajectory the field will track the instantaneous minimum of the potential. This is indeed the case, since the radius of curvature is large $(\sim M_{\text {pl}})$, and then the angular velocity $\dot{\theta}$ -which sets the strength of the centrifugal force that displaces the field from the instantaneous minimum- is small during inflation\footnote{If the radius of curvature is small (and then a monodromy in $\theta$ must be assumed to have large field inflation) then the effects of the displacement from the radial minimum may become very important \cite{Achucarro:2015rfa}.}. If the observable scales leave the horizon when the heavy radial field is stabilized in its minimum, the predictions for the inflationary observables will coincide with the single field predictions of (\ref{eq:potvai}). There is still the possibility that there 
is a first 
period of inflation driven by the heavy radial field, \emph{\`{a} la} chaotic inflation, and if the observable scales leave the horizon during or before the transition this may cause departures from the single field predictions of (\ref{eq:potvai}). We compute the predictions for this particular case in the last section. 
The more interesting case is for $\beta\gtrsim 1$, for which the mass eigenvalues are of the same order in some part of the trajectory. Furthermore, for $\beta>1.5$ there is no minimum in the radial direction for small values of $\theta$. This will ensure that the dynamics are of a multi-field nature. As the hierarchy is restored when approaching the absolute minimum at $\theta=\pi$, the isocurvature perturbations will be eventually damped. We will study this case in section IV. Let us note that the presence or absence of a hierarchy between the masses does not ensure that inflation will happen, as inflation requires at least one of the eigenvalues to be smaller than the Hubble parameter. By numerically evaluating the background we will see that many trajectories have indeed enough e-foldings of inflation.

In any multifield model one should be concerned about the presence of isocurvature modes in the CMB, as they are heavily constrained by Planck data \cite{Ade:2015lrj}. The constraints are usually stated in terms of the primordial isocurvature fraction, defined as
\be
\beta_{iso}(k)=\frac{P_{\mathcal{I} \mathcal{I}}(k)}{P_{\RR \RR}(k)+P_{\mathcal{I} \mathcal{I}}(k)} \ ,
\ee
where $P_{\RR\RR}(k)$ and $P_{II}(k)$ are the power spectra of the curvature and isocurvature perturbations. The Planck constraints are given specifically for a variety of models, in which the isocurvature component is attributed to one of the different elements of the plasma, and different correlations are assumed between the curvature and isocurvature components. None of these idealized situations corresponds exactly to our case as we do not specify, for example, a mechanism for reheating.  We can however take a nominal value of $\beta_{iso}<0.01$, which is the typical order of magnitude for these constraints.

While the isocurvature mode may be relevant when the observable scales left the horizon, from the time of inflation to the decoupling of the CMB there is ample time and a diversity of physical processes. There are indeed many ways in which isocurvature perturbations during inflation may decay so that at the time of decoupling we only happen to measure adiabatic fluctuations. One possibility is that thermal equilibrium is achieved in the plasma era \cite{Weinberg:2004kf}.  Another possibility is that the field with isocurvature perturbations becomes heavy between horizon crossing and the end of inflation (see, e.g \cite{Meyers:2010rg,Turzynski:2014tza,Ellis:2014opa,Bielleman:2015lka}). One can see from figure \ref{fig:eig} that when the absolute minimum of the potential is reached (at $\theta=\pi$), the hierarchy between the eigenvalues is restored ($\lambda_2\gg \lambda_1$). If the trajectories follow this minimum before the end of inflation, the isocurvature mode will be rapidly damped, and the model will 
be consistent with the Planck 
isocurvature bound. 

Another consequence of the presence of an additional dynamical direction in the axion potential has to do with the initial conditions. In the single field potential (\ref{eq:potvai}), a number $N_{\star}$ of e-folds are achieved by setting the initial angle at $\theta_{\star}$ given by
\be
\theta_{\star}=2\arcsin\[\(1+\frac{1}{2f^2}\)^{1/2}e^{-N_{\star}/2f^2}\] \ .
\ee
It is then possible to assign a probability for having more than $N_\star$ e-folds by assuming, e.g. a linear probability distribution for $\theta_{\text{ini}}$ in the interval $[0,2\pi]$. For example, for $f=1 M_{\text{pl}}$ there is a 20\% probability of having more than 60 e-folds (ignoring the fact that the patches of the Universe that inflate are much bigger than those who don't, and that this enhances the probability of observing those patches). When considering the effects of the radial direction in field space, we will find that it is possible to inflate for values of $\theta_{\text{ini}}<\theta_{\star}$.    

 \section{Equations of Motion}
In order to compute the predictions for the model we first solve the background equations of motion. For a general multifield model with $n$-fields $\phi(t)^a$ ($a$ ranging from 1 to $n$), and field space metric $\gamma_{a b}$, the background equations  of motion are\footnote{In this section, we set the reduced Planck mass $m_{\text{pl}}=M_{\text{pl}}/8\pi^2$ to 1.}
\ba\label{eq:eqs_mf}
&&D_t \dot{\phi}_0^a  +3H\dot{\phi}^a_0+V^a=0 \, \\
&&3 H^2 = \dot{\phi_0^2}/2 +V, 
\ea

\noindent where $\phi_0(t)^a$ is the time dependent background component of the field $\phi^a(t)$, $\dot \phi_0^2 \equiv \gamma_{a b} \dot \phi_0^a \dot \phi_0^b$, $H= \dot{a} / a$ and $D_t X^a  = \dot X^a + \Gamma^{a}_{b c} \dot \phi_0^b X^c $ is a covariant time derivative, with
\be
\Gamma^{a}_{b c} =  \gamma^{a d} (\partial_b \gamma_{d c}  + \partial_c \gamma_{b d} - \partial_d \gamma_{b c} )/2\ .
\ee

In our case, working in polar coordinates such that $\phi^a(t)=(r,\theta)$ and with flat field space metric $\gamma_{ab}$ ($\gamma_{11}=1$,$\gamma_{22}=r^2$ and $\gamma_{12}=\gamma_{21}=0$), the equations of motion reduce to:
\ba
&&\ddot{r}+3H\dot{r}-r\dot{\theta}^2+\frac{m^4}{r_0}\[-2\frac{r}{r_0}\(1-(\frac{r}{r_0})^2\)+\beta\cos \theta\] =0 \ ,\\
&&r^2\ddot{\theta}+2r\dot{r}\dot{\theta}+3Hr^2\dot{\theta}-m^4\beta\frac{r}{r_0}\sin \theta=0 \ ,
\ea
and
\be
3H^2=\frac{1}{2}\dot{r}^2+\frac{1}{2}r^2\dot{\theta}^2+m^4\[\(1-\(\frac{r}{r_0}\)^2\)^2+\beta\(\frac{r}{r_0}\)\cos \theta\]\ .
\ee
Having solved the background trajectory, it is useful to define vectors parallel and perpendicular to the trajectory. This set of vectors forms a basis on which the equations for the perturbations can be projected. The curvature perturbations are nothing more than the projection along the tangential direction, while the isocurvature perturbation is the projection on the perpendicular direction~\cite{Gordon:2000hv,GrootNibbelink:2000vx,Achucarro:2010da}. The tangential vector is given by $T^a=\dot{\phi^a}/\dot{\phi_0}$, and the normal vector is constructed such that $T_aN^a=0$ and $N_aN^a=1$ (indices are raised and lowered with the field space metric $\gamma_{ab}$). In the two-field case it is given by $N_a=(\text{det} \gamma)^{1/2}\epsilon_{ab}T^b$ where $\epsilon_{ab}$ is the two dimensional Levi-Civita symbol with $\epsilon_{11}=\epsilon_{22}=0$ and $\epsilon_{12}=-\epsilon_{21}=1$, such that
\be
T^a=\frac{1}{\sqrt{\dot{r}^2+r^2\dot{\theta}^2}}\big( \dot{r} , \dot{\theta} \big) \qquad \text{and} \qquad N^a=\frac{r}{\sqrt{\dot{r}^2+r^2\dot{\theta}^2}}\(\dot{\theta} ,  -r^{-2}\dot{r} \) \ . 
 \ee
The rate of change of the tangential vector defines the angular velocity of the trajectory $\dot{\theta}$, as $D_tT^a\equiv-\dot{\theta}N^a$. The slow-roll parameters can be written as
\be
\epsilon \equiv -\dot{H}/H^2 \quad,\quad \eta^a\equiv-\frac{1}{H\dot{\phi_0}} D_t\dot{\phi}_0^a \ .
\ee
While the slow-roll parameter $\epsilon$ is a scalar, the change in the inflaton velocity is a two dimensional vector. We may decompose $\eta^a$  along the normal and tangent directions by introducing two independent parameters $\eta_{\parallel}$ and $\eta_{\perp}$ as
\be
\eta^a=\eta_{\parallel}T^a+\eta_{\perp}N^a \ .
\ee
Then, one finds that
\ba
\eta_{\parallel}=- \frac{\ddot{\phi_0}}{H\dot{\phi_0}} \ , \\
\eta_{\perp}=- \frac{V_N}{H\dot{\phi_0}} \label{eq:eta_perp_eq} \ .
\ea
where $V_N=N^a\partial_a V$. It is easy to see that $\eta_{\perp}$ is just the angular velocity in e-folds, $\eta_{\perp}=\dot{\theta}/H$, and that sufficient inflation only demands $\epsilon$ and $\eta_{\parallel}$ to be small. In flat gauge, the curvature and isocurvature perturbations are given by~\cite{Achucarro:2012sm}
\be
\mathcal{R}\equiv-\frac{H}{\dot{\phi^a}} T_a\delta\phi^a \hspace{1cm}\text{and} \hspace{1cm} \mathcal{F}=N_a\delta\phi^a \ , 
\ee
which satisfies the following equations of motions
\ba
\label{eq-of-motion-R}
\ddot {\mathcal R} +  (3  + 2 \epsilon - 2 \eta_{||}) H \dot  {\mathcal R} + \frac{k^2  }{a^2}  {\mathcal R} & = &  2  \dot \theta  \frac{H}{\dot \phi_0} \left[ \dot {\mathcal F} + \bigg( 3 - \eta_{||}  - \epsilon + \frac{\ddot\theta}{H\dot\theta} \bigg) H {\mathcal F} \right] \, ,  \\
\label{eq-of-motion-F}
\ddot {\mathcal F} +  3  H \dot  {\mathcal F} + \frac{k^2 }{a^2}  {\mathcal F}  + M_{\rm eff}^2  {\mathcal F} & = & - 2 \dot \theta \frac{\dot \phi_0}{H}  \dot {\mathcal R}  \, .
\ea
where $M_\text{eff}^2=m^2-\dot{\theta}^2$, and $m^2\equiv N^aN^bV_{ab}$ is the bare mass of the field ${\mathcal F}$. Imposing the Bunch Davies initial conditions when the modes are well inside the horizon, we compute 
the predictions for different initial positions in field space, and compare them  with the Planck confidence contours in the ($n_s,r$) plane and bounds on isocurvature fluctuations. We repeat this analysis for different values of $r_0$ and $\beta$.\\

\section{Trajectories with no mass hierarchy}

In the case in which $\beta \sim 1$, which we will study in the following sections, both masses are comparable and we are completely in a two-field regime

\subsection{Case 1: $r_0=1 M_{\text{pl}}$}

\begin{figure}[ht!]
  \includegraphics[width=220pt]{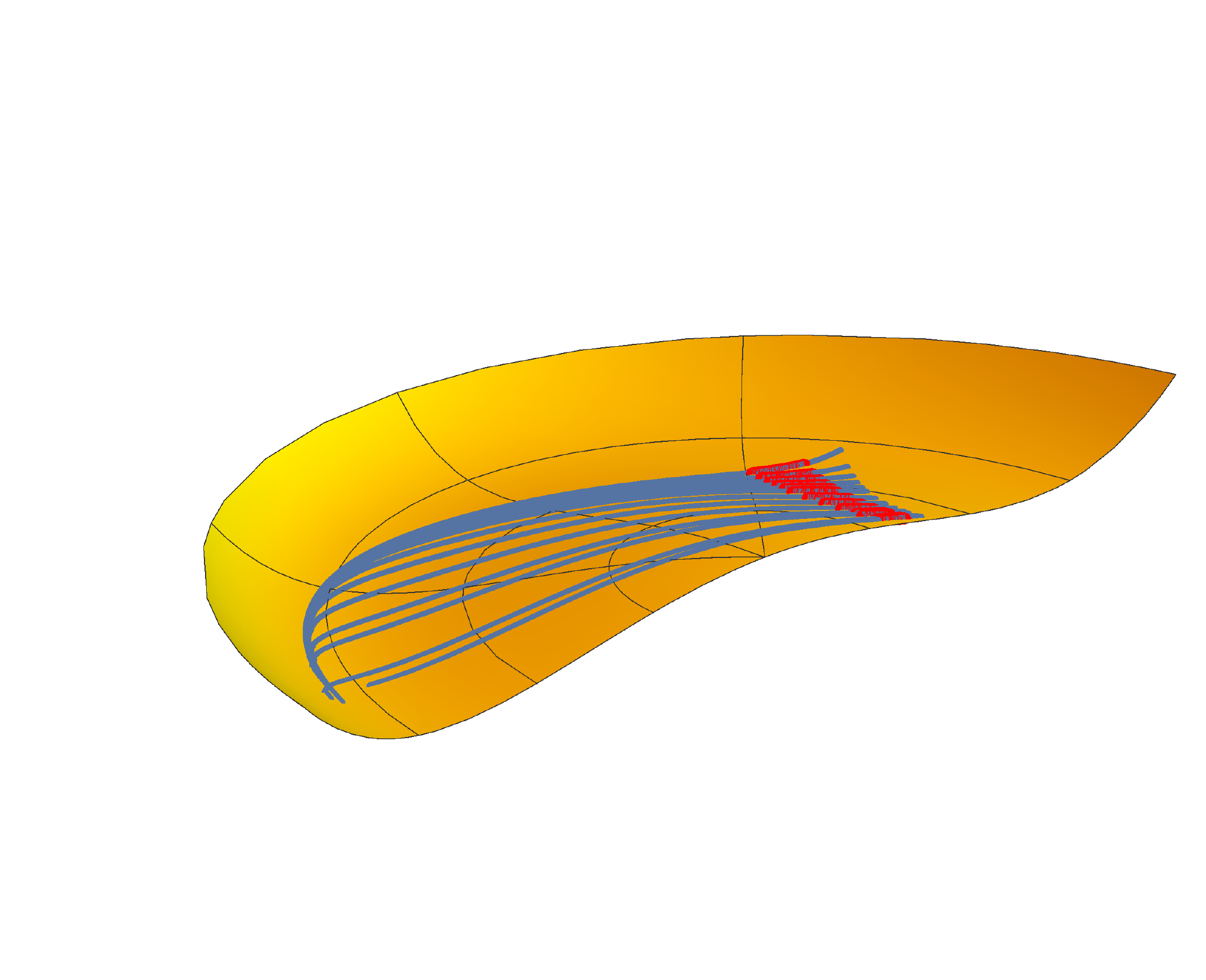}
\includegraphics[width=240pt]{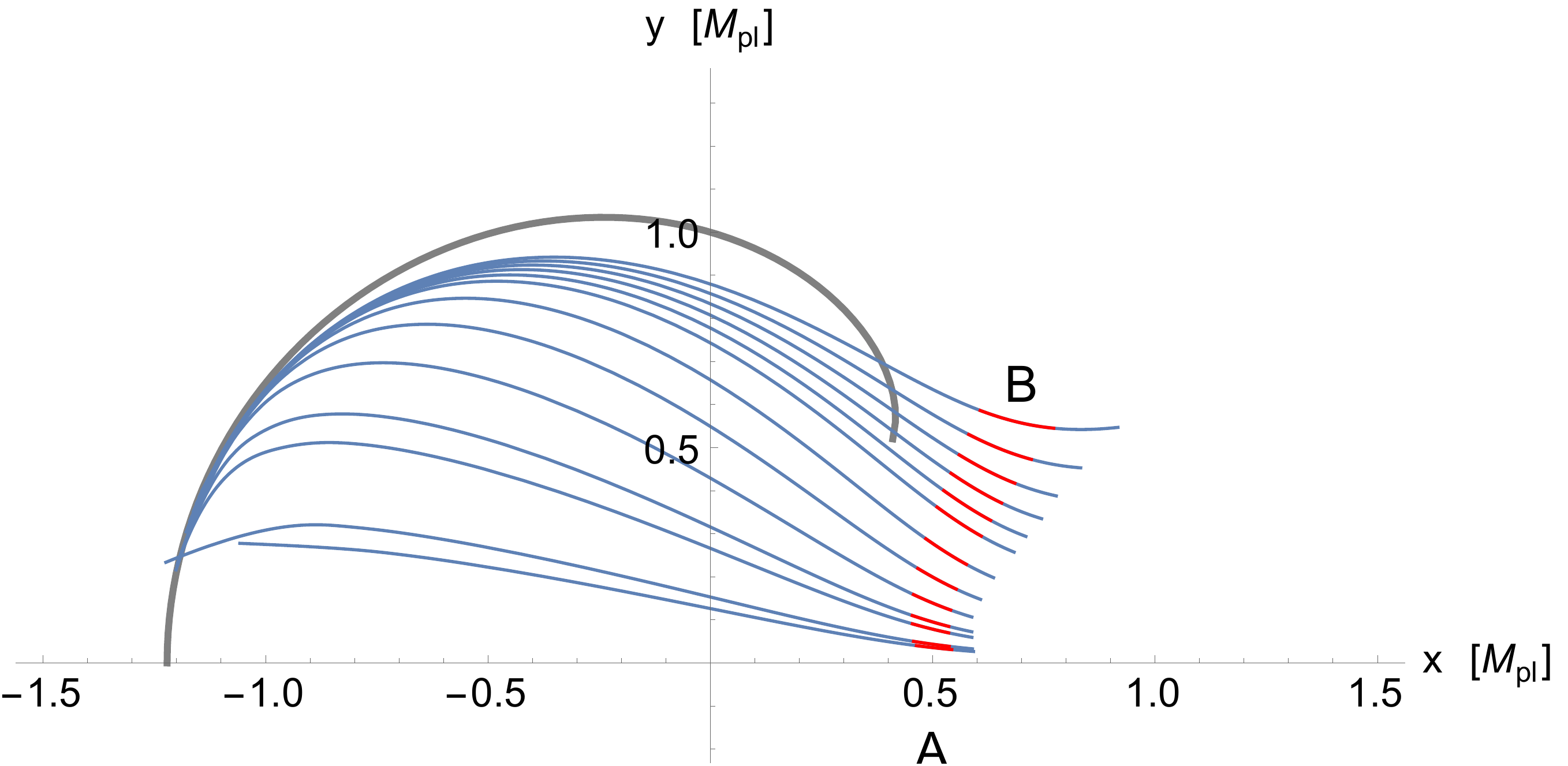}
  \caption{Different trajectories for different initial conditions in the case $\beta=2.4$, where $x=r\cos\theta$ and $y=r\sin\theta$. We indicate in red where the pivot scale left the horizon (50 to 60 e-folds before the end of inflation). The gray line tracks the instantaneous minimum of the potential in the radial direction.}
\label{fig:traj1}
\end{figure}

An important threshold value for the axion decay constant is the Planck mass. While $f \leqslant 1 M_{\text{pl}}$ is in tension with the data, it seems difficult to achieve  $f \geqslant 1 M_{\text{pl}}$ in well controlled models. As we have seen, for values of $\beta\sim \OO(1)$ the radial field is not massive enough to make the trajectories follow their instantaneous minimum. We have to numerically evolve the equations of motion in order to compute the trajectories. For a nominal value of $\beta=2.4$, these trajectories can be seen in figure \ref{fig:traj1}.

\begin{figure}[ht!]
  \includegraphics[width=200pt]{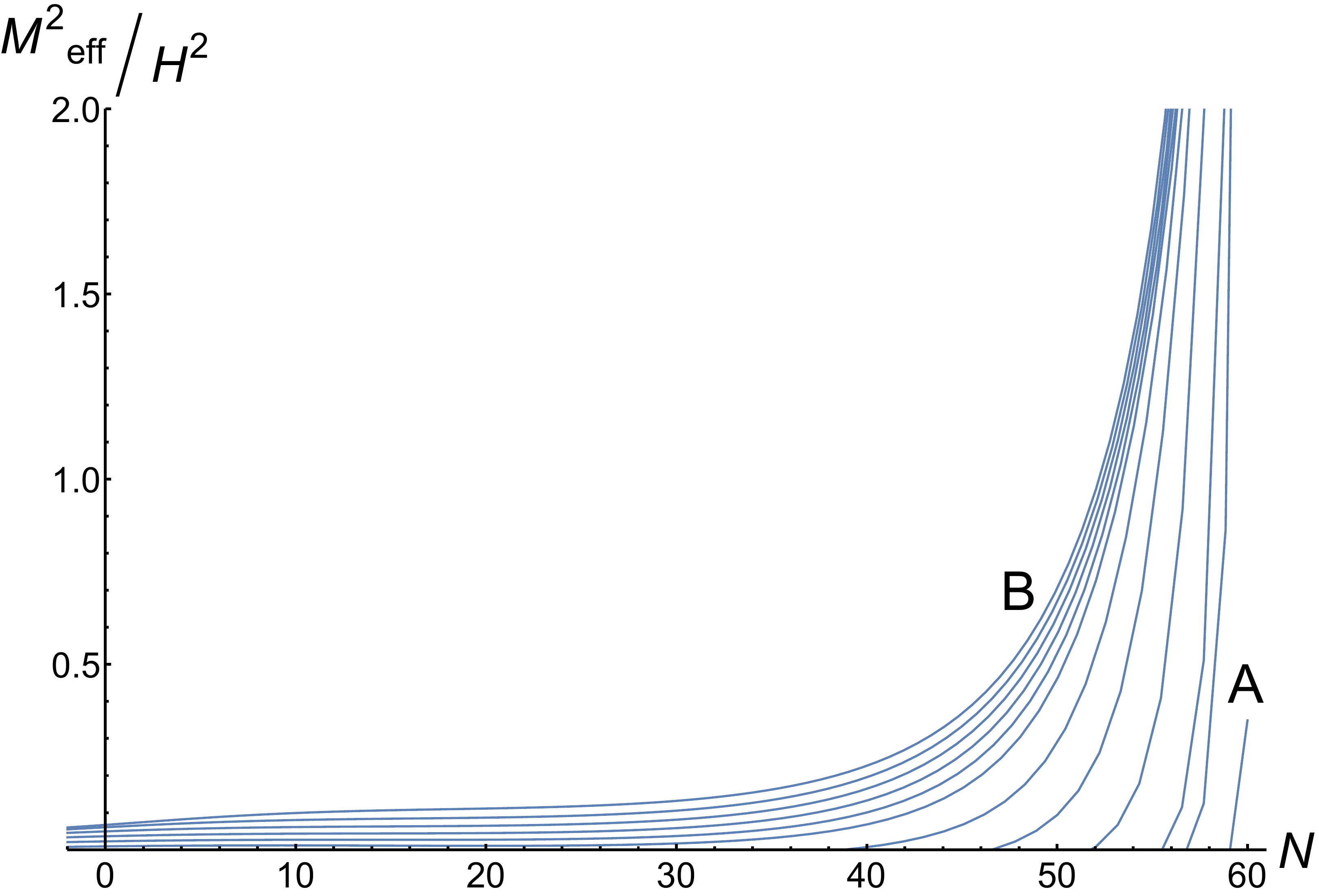}
  \includegraphics[width=200pt]{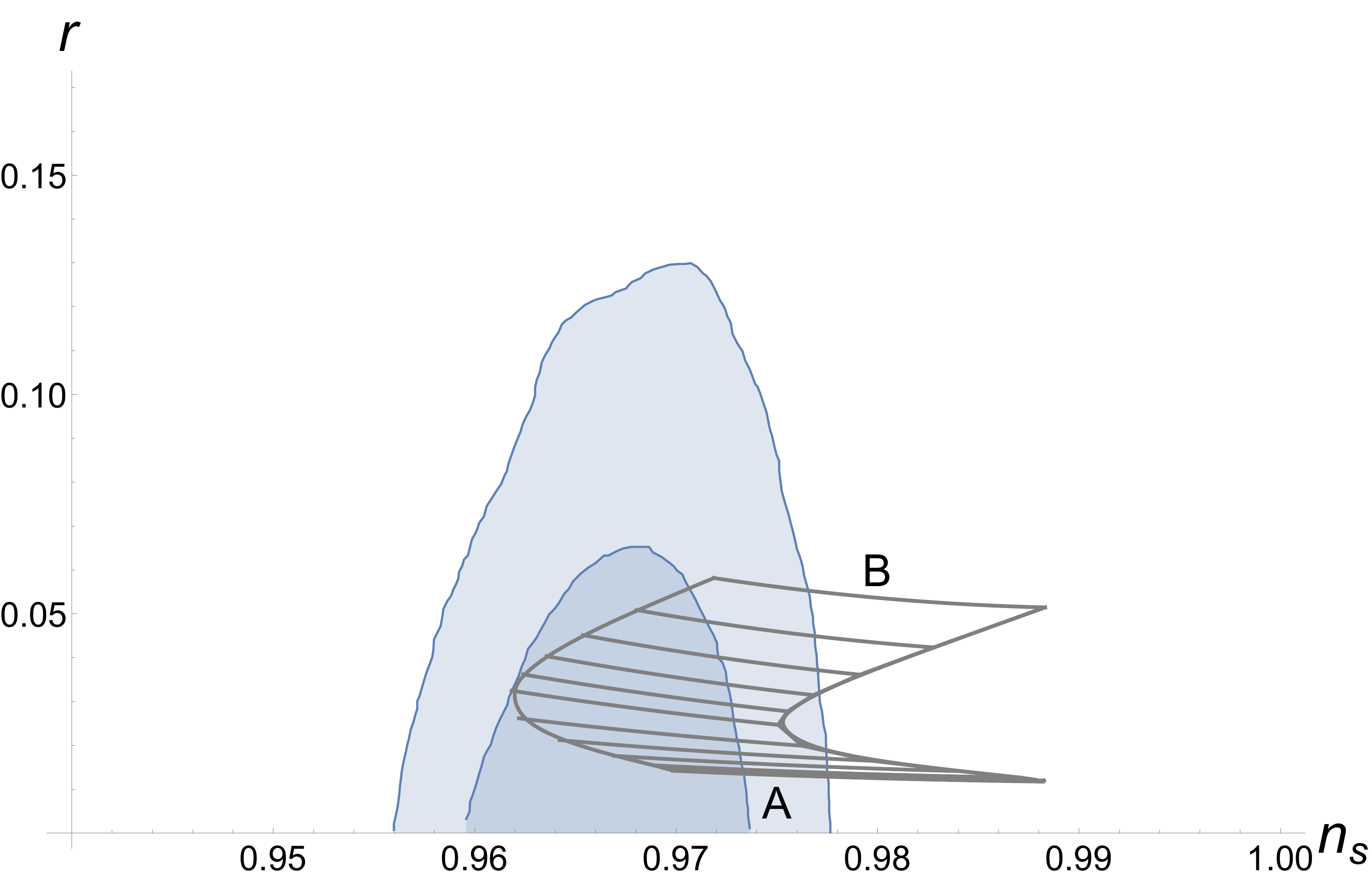} 
\caption{\emph{Left}: The orthogonal mass to the trajectories for different initial conditions in the case $\beta=2.4$, as a function of number of e-folds N (N=60 is the end of inflation). \emph{Right}: The $(n_s,r)$ plane for the different initial conditions. The width represents the predictions for 50 to 60 e-folds before the end of inflation, and the shaded regions are the 1 and 2-$\sigma$ confidence contours as given by Planck.}
\label{fig:mrns1}
\end{figure}

\begin{figure}[ht!]
  \includegraphics[width=500pt]{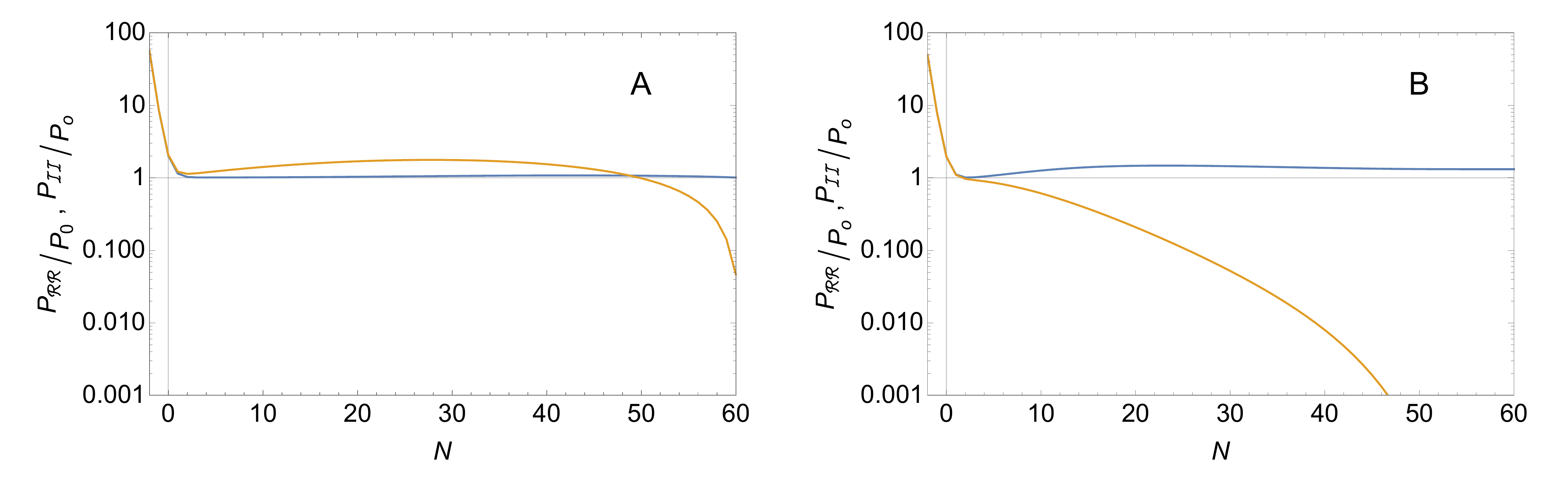} 
\caption{Curvature (blue) and isocurvature (yellow) perturbations for $\beta=2.4$ and $r_0= M_{\text{pl}}$, normalized to $P_0\equiv H^2/8\pi^2\epsilon$ (with $H$ and $\epsilon$ evaluated at horizon crossing). The left panel is for trajectory $A$, in which the isocurvature perturbations do not have time to decay. The right panel is for trajectory $B$, in which the isocurvature perturbations decay at the end of inflation.}
\label{fig:curv_iso_1}
\end{figure}

As one can see, the observable scales are placed at different values of $r,\theta$ for different initial conditions, therefore the predictions of the model will depend upon them. Additionally, nearly all of the trajectories converge at the end of inflation. This happens because the mass in the orthogonal direction is increasing, as can be seen in figure \ref{fig:mrns1}.  While the orthogonal mass is sub-Hubble for most  of the trajectory -which ensures the two-field effect to be important- it becomes super-Hubble for around 10 e-folds before the end of inflation. This will make the isocurvature perturbations suppressed and unobservable. There are however some trajectories, close to trajectory A in figure \ref{fig:traj1}, in which the minimum is never reached before the end of inflation. We expect those trajectories to have a non-negligible amount of isocurvature fluctuations.
In figure \ref{fig:curv_iso_1} we plot the amount of curvature and isocurvature perturbations (normalized by the single field prediction $P_0\equiv H^2/8\pi^2\epsilon$, evaluated at horizon crossing) as a function of the number of e-folds $N$ for $k_{60}$, the mode that left the horizon 60 e-folds before the end of inflation. We can see the isocurvature perturbations decay significantly in the last 10 e-folds of inflation for trajectories close to trajectory B -such that adiabaticity is reached- while they remain large for trajectories close to A.
We can plot the amount of isocurvature perturbations as a function of the trajectory, which we choose to parametrize by $\theta_{60}$, the angle of the trajectory 60 e-folds before the end of inflation. 
\begin{figure}[ht!]
  \includegraphics[width=0.6\textwidth]{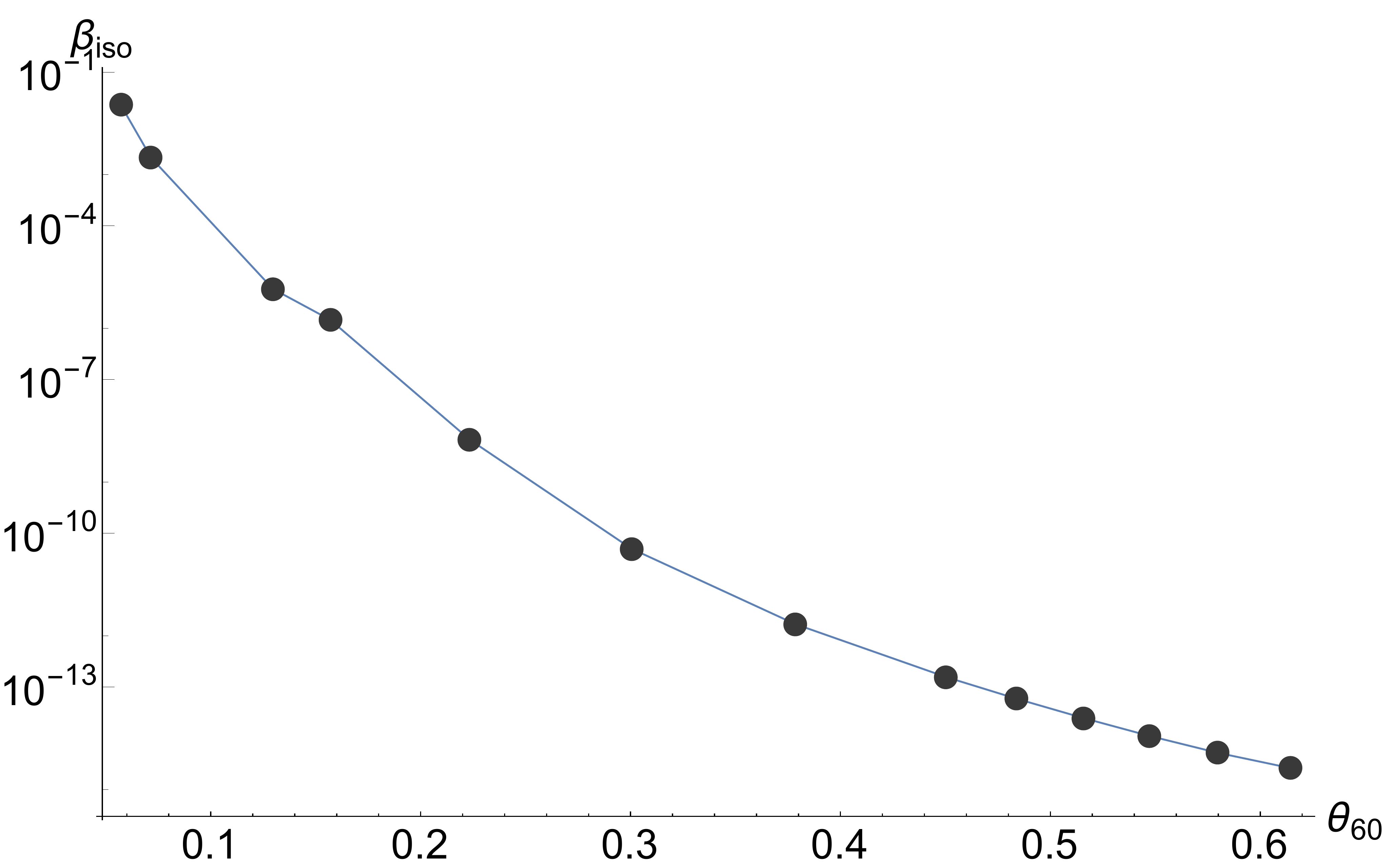} 
\caption{Isocurvature perturbations at the end of inflation for $\beta=2.4$ and $r_0= M_{\text{pl}}$ as a function of $\theta_{60}$, the angle of the trajectory 60 e-folds before the end of inflation. Trajectories with smaller $\theta_{60}$ have larger amounts of isocurvature perturbation since they experience the mass hierarchy for a shorter time.}
\label{fig:bvstf1}
\end{figure}

From figure \ref{fig:bvstf1} we can see that only the trajectories with $\theta_{60}\ll 1$ will have unsuppressed isocurvature fluctuations. The only way to know whether this is a big or small subspace is with a theory of initial conditions, which we do not provide here.

For these values of the parameters ($r_0=M_{\text{pl}}$ and $\beta=2.4$) we find that $m^4\sim10^{-10}$ $M_{\text{pl}}^4$ is needed in order to fix the amplitude of the primordial curvature perturbations ($H^2/8\pi^2\epsilon\sim 10^{-9}$). This translates into $\Lambda/r_0\sim 10^{-4}$. This is the \emph{same} order of magnitude needed in the single field natural potential: a smaller value of $\beta$ (e.g. 0.01) needs a larger value of $m^4(\sim10^{-8} $ $M_{\text{pl}}^4$) to generate the right amplitude for the fluctuations, resulting in $\Lambda/r_0$ of the same order.


\subsection{Case II: $r_0=0.8 M_{\text{pl}}$}

\begin{figure}[ht!]
  \includegraphics[width=240pt]{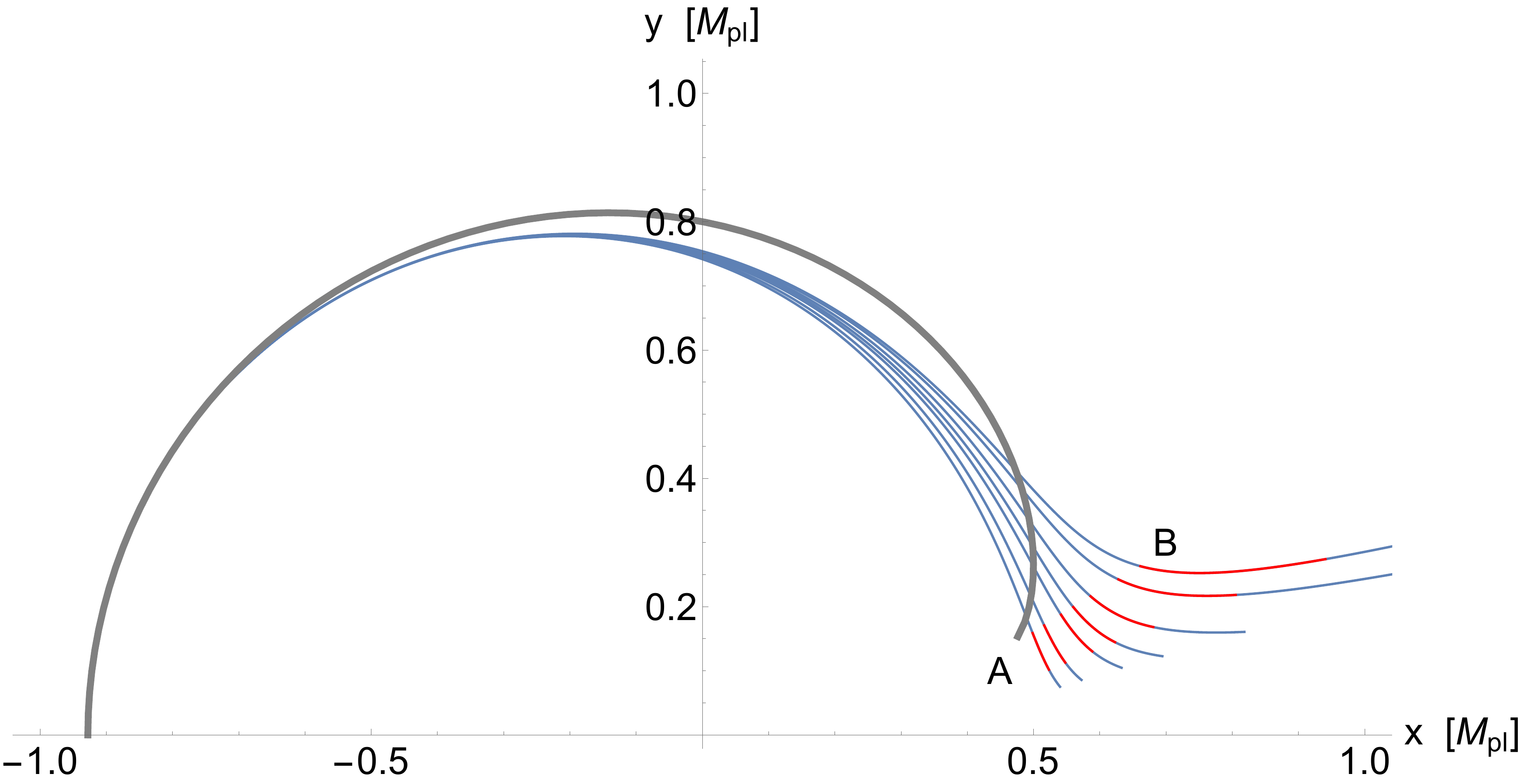}
   \includegraphics[width=200pt]{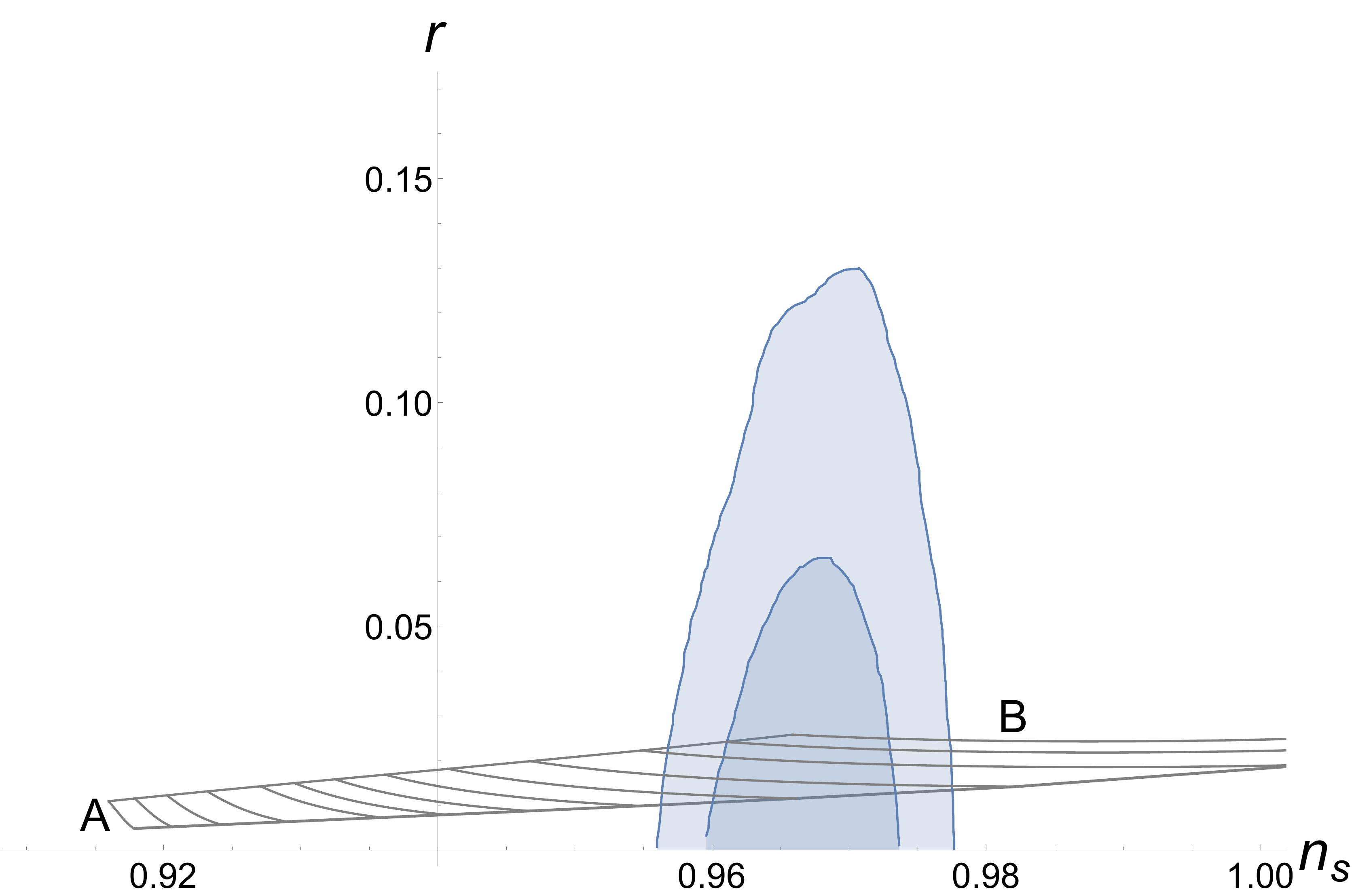} 
\caption{\emph{Left}: Different trajectories for different initial conditions in the case $\beta=1.6$. In red we show where the pivot scale left the horizon (50 to 60 e-folds before the end of inflation). The gray line is the minimum of the potential in the radial direction. \emph{Right}: Predictions in the $(n_s,r)$ plane for the depicted trajectories. The width represents the predictions for 50 to 60 e-folds before the end of inflation.}
\label{fig:traj2}
\end{figure}

As subplanckian values of the axion decay constant may be better understood in terms of an effective field theory, it is interesting to know whether the predictions for such values are also in better agreement with the data. We take $r_0=0.8 M_{\text{pl}}$ as a test case. For comparison, in this case we choose a slightly smaller $\beta=1.6$.  We will expect then the trajectories to be more confined to the instantaneous minimum of the potential. We show the trajectories and their predictions in the $(n_s,r)$ plane for different initial conditions in figure \ref{fig:traj2}. 
  
\begin{figure}[ht!]
  \includegraphics[width=500pt]{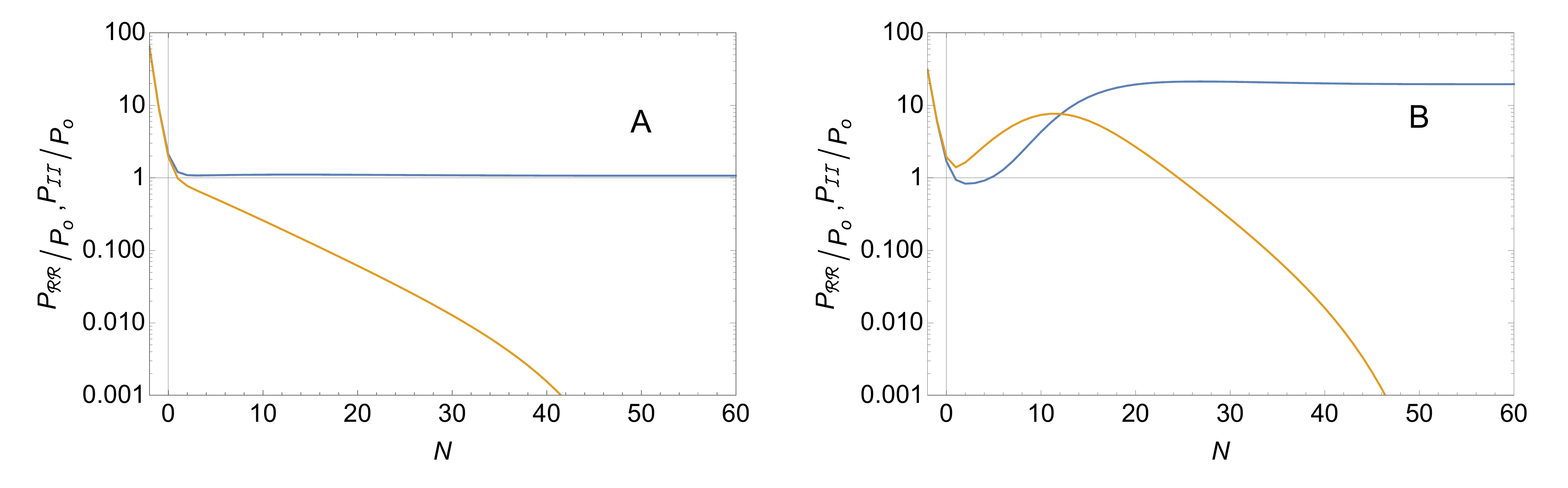}
\caption{Curvature (blue) and isocurvature (yellow) perturbations for $\beta=1.6$ and $r_0=0.8 M_{\text{pl}}$. normalized to $P_0\equiv H^2/8\pi^2\epsilon$ (with $H$ and $\epsilon$ evaluated at horizon crossing).  For both trajectories $A$ and $B$, left and right panel respectively, the isocurvature mode has decayed at the end of inflation. For trajectory B the amplitude of the curvature mode is severely affected by isocurvature perturbations.}
\label{fig:iso2}
\end{figure}

The amount of isocurvature perturbations for trajectories A and B are shown in figure \ref{fig:iso2}.

As all the trajectories reach the attractor several e-folds before the end of inflation, the isocurvature perturbations are strongly damped. A larger fraction of isocurvature perturbations can be found by increasing the value of $\beta$. We can also see that, for trajectories close to B, the sourcing of isocurvature to curvature fluctuations has a big impact on the latter. In this case the amplitude is enhanced by $\sim 10$ times compared to their value at horizon crossing. Additionally the running of the spectral index is enhanced.

\section{Trajectories with mass hierarchy}

\begin{figure}[ht!]
  \includegraphics[width=180pt]{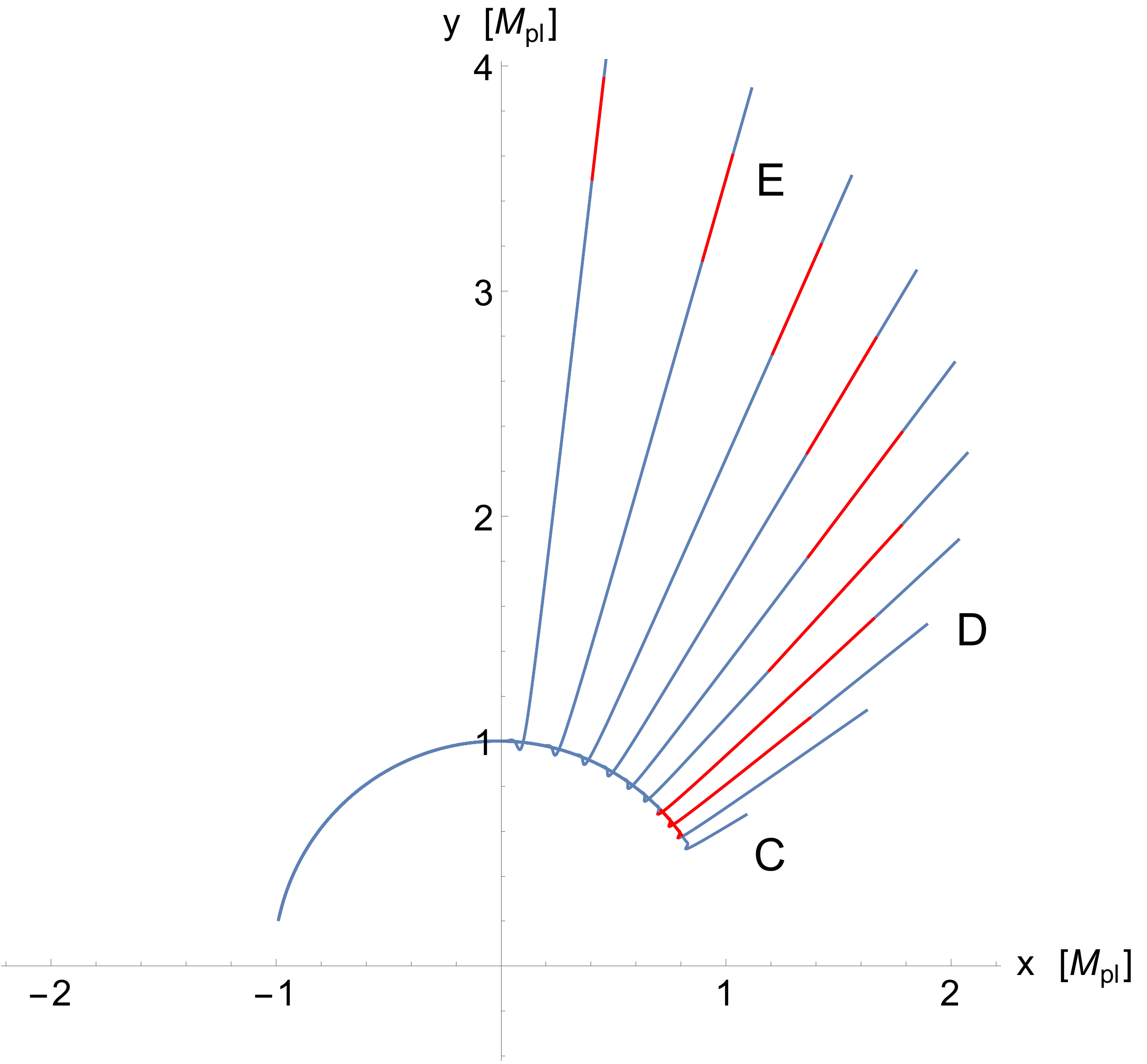}   
\caption{Different trajectories for different initial conditions in the case $\beta=0.08$. In red is where the observable scales left the horizon (50 to 60 e-folds before the end of inflation).} 
\label{fig:traj_beta_001}
\end{figure}

For completeness, we also compute the predictions in the case in which the U(1) symmetry is mildly broken. In this case $\beta\ll 1$ and the hierarchy of masses between the radial and angular field is large throughout the trajectory. Let us note that if the field is at a large radius, the system will also inflate, \emph{\`{a} la} chaotic inflation. The dynamics admits then three distinct situations. 
\begin{itemize}
\item Inflation starts either in $r$ or $\theta$, ends in $\theta$, and the observable e-folds left when the inflaton was $\theta$ (e.g. trajectory C in fig. \ref{fig:traj_beta_001})
\item There is a first period of inflation in the radial direction, a second period in $\theta$, and the observable e-folds left during the transition. (e.g. trajectory D in fig. \ref{fig:traj_beta_001} )
\item There is a first period of inflation in the radial direction, a second period in $\theta$, and the observable e-folds left when the inflaton was $r$. Both periods of inflation could be matched or not. (e.g. trajectory E in fig. \ref{fig:traj_beta_001} )
\end{itemize}

\begin{figure}[ht!]
  \includegraphics[width=220pt]{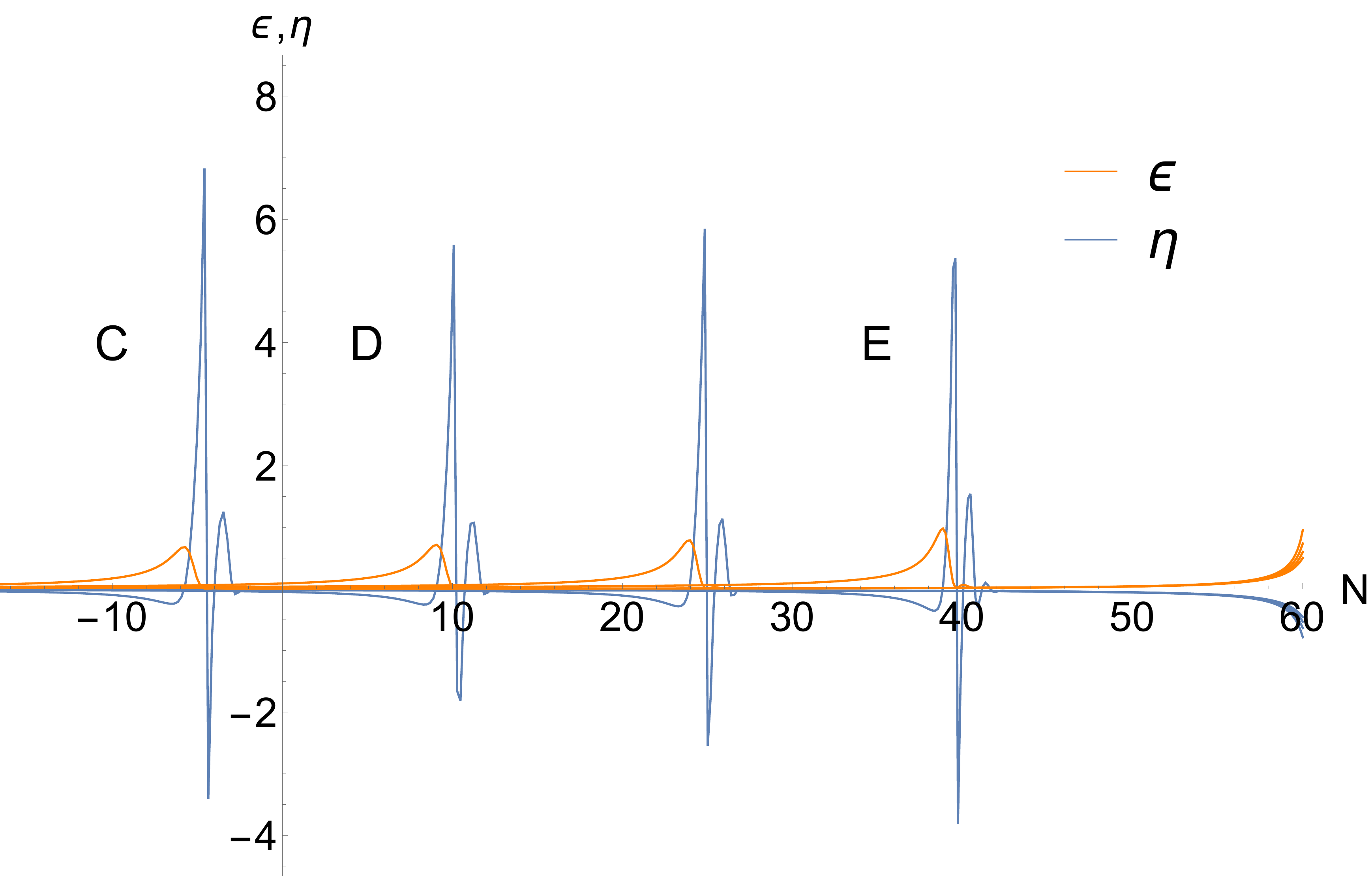}
 \includegraphics[width=220pt]{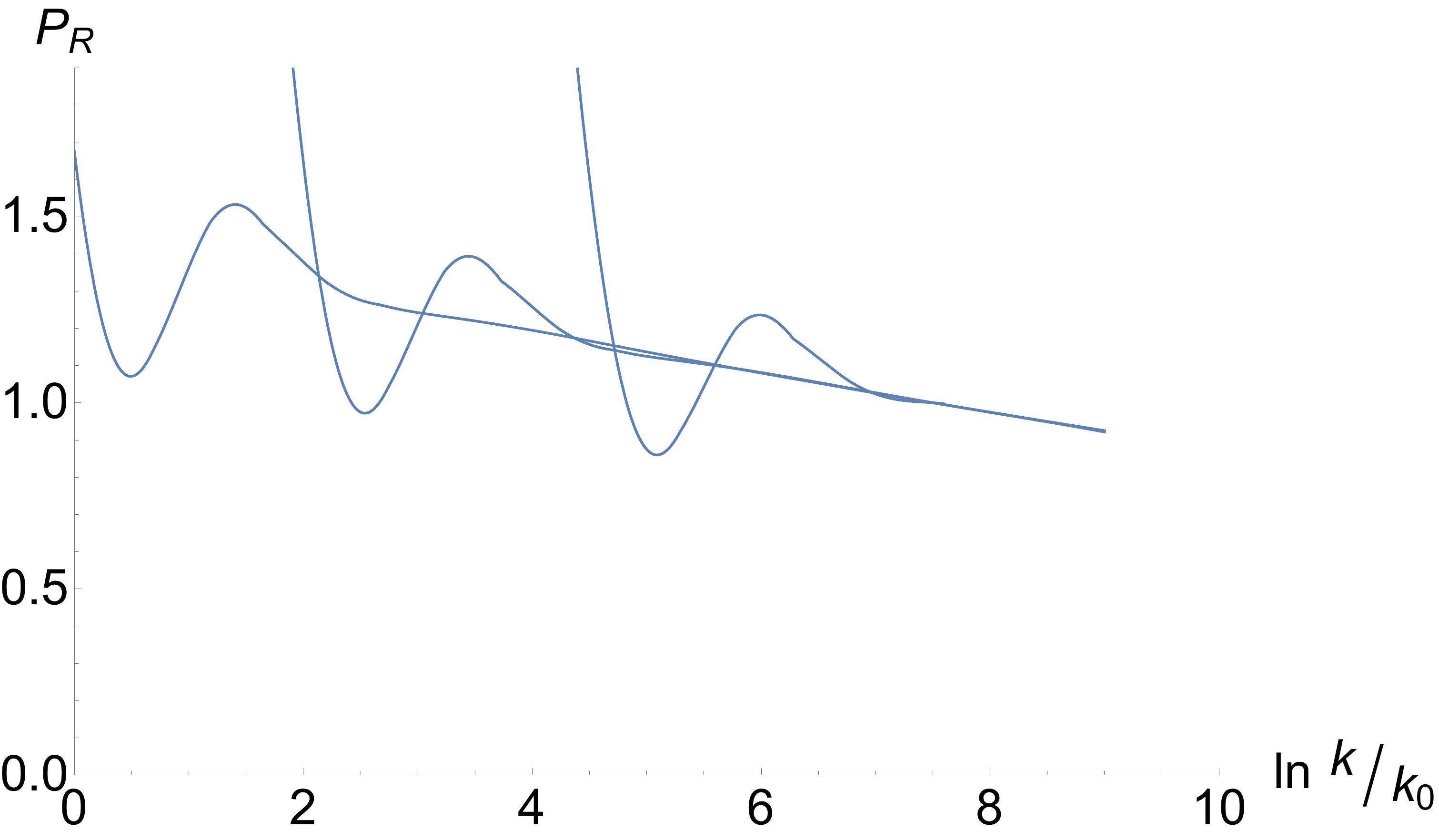}
  \caption{\emph{Left}: The slow parameters $\epsilon$ and $\eta$ ($\equiv \dot{\epsilon}/H\epsilon$) as a function of e-folds (N=60 is the end of inflation). 
 \emph{Right}: Power spectrum for different trajectories in the case $\beta=0.08$, as a function of $\ln (k/k_0)$, with $k_0$ the scale that left the horizon 60 e-folds before the end of inflation.}
\label{fig:Preds_beta_001}
\end{figure}

The transition from inflating in $r$ to inflating in $\theta$ will produce a local peak in the slow-roll parameters, as can be seen in figure \ref{fig:Preds_beta_001}. The predictions for these cases are different. The first case will yield the same predictions as the single natural inflation potential. The second and third are more interesting. 
For trajectories like $D$ and $E$ and we expect large values for the slow-roll parameters at horizon crossing. If it were not for the additional e-folding provided by $\theta$, those scales would correspond to very few e-folds before the end of inflation, in which the slow-roll parameters are big. This will also imply that the predictions for the primordial power spectrum may largely differ from a power law. In single field inflation, a period of fast roll at the time when the largest observable scales left the horizon might be an interesting mechanism to generate a smaller amplitude for the curvature perturbations at those scales \cite{Contaldi:2003zv,Cicoli:2013oba,Pedro:2013pba,Bousso:2013uia}, as suggested by observations (most recently, Planck 2015 results\cite{Ade:2015lrj}). As explained in \cite{Blanco-Pillado:2015bha}, this does not generalize to multifield models, as curvature perturbations may grow after horizon crossing. We find that this is the case in our model, such that perturbations on large 
scales are in general bigger than at smaller scales. Now, because of the small oscillatory feature in the slow-roll parameters, there is a zone in which the perturbations become smaller than the flat power spectrum. In figure \ref{fig:Preds_beta_001} we show the power spectrum in the cases in which the predictions do not resemble a power law in all the range of observable e-folds, which is the case for trajectories that are subject to a turn in these scales. The rest of the trajectories are ruled out by direct computation of $(n_s,r)$. 

\vfill

\section{Conclusions}

Pseudo Nambu-Goldstone bosons are interesting candidates for driving inflation.  However, the single field description with a sinusoidal potential leads to predictions that are in tension with the most recent CMB data.  In this paper, we have shown that the simplest two-field completion of natural inflation (the original model proposed by Freese et al. in \cite{Freese:1990rb}),  has a regime in which its predictions are consistent with observations.  To do so, we have considered the possibility that the mass of the angular field (an axion) is of the same order as its radial partner.  The normalization of the two-point function then fixes a hierarchy between the scales of spontaneous and explicit symmetry breaking, thus keeping higher order corrections under control.  Isocurvature perturbations,  while important for sourcing the curvature perturbations around the time of horizon crossing,  decay before the end of inflation, since a mass hierarchy is created at the end of the inflationary trajectory.  This makes the model also compatible with Planck isocurvature bounds.

For completeness, we have also computed the predictions for the more standard regime in  which  the  radial  field  is  very  massive,  but  in  the  case  in  which  the  initial  conditions are such that there are two stages of inflation, first in the radial direction and then in the angular direction. We find that  in  general  this  will  imply  an  initial  period  of  fast  roll, which in this particular multifield setting (and contrary to single field models) provides an enhancement of power on large scales. \\

\textbf{Acknowledgements} We would like to thank Yvette Welling for many interesting discussions, and for sharing with us her multifield code. This work was supported by a Leiden Huygens Fellowship (VA), the Netherlands Foundation for Fundamental Research on Matter (F.O.M.) under the program “Observing the Big Bang” (AA), the Basque Government grant IT559-10, the Spanish Ministry of Science and Technology grant FPA2012-34456, the Spanish Consolider-Ingenio 2010 program CPAN CDS2007-00042 (AA), by Grants-in-Aid for Scientific Research from the Ministry of Education, Culture, Sports, Science, and Technology (MEXT), Japan, No. 25400248(MK), No. 23104008 (FT), JSPS KAKENHI Grant Numbers 24740135 (FT), 26247042(FT), and 26287039 (FT), MEXT Grant-in-Aid for Scientific Research on Innovative Areas No.15H05889 (MK and FT) and by the World Premier International Research Center Initiative (WPI), MEXT, Japan (MK and FT).



\bibliography{2fieldnat}

\end{document}